\documentclass[aps,prl,reprint,preprintnumbers,floatfix,superscriptaddress]{revtex4-1}
\usepackage{graphicx}
\usepackage{lineno}
\usepackage{subfigure}
\usepackage[colorinlistoftodos]{todonotes}
\usepackage{color}
\usepackage{xspace}
\usepackage{siunitx}

\graphicspath{{figures/}}

\newcommand{\nue}{\ensuremath{\nu_{e}}\xspace}

\newcommand{\nuebar}{\ensuremath{\bar\nu_{e}}\xspace}
\newcommand{\antinue}{\nuebar}

\newcommand{\numu}{\ensuremath{\nu_{\mu}}\xspace}

\newcommand{\numubar}{\ensuremath{\bar\nu_{\mu}}\xspace}
\newcommand{\antinumu}{\numubar}
\newcommand{\nutau}{\ensuremath{\nu_{\tau}}\xspace}

\newcommand{\LL}{\ensuremath{-2\ln \cal{L}}\xspace}

\newcommand{\dmsq}{\ensuremath{\Delta m^2_{32}}\xspace}
\newcommand{\snsq}{\ensuremath{\sin^2 \theta_{23}}\xspace}
\newcommand{\deltacp}{\ensuremath{\delta_{\rm{CP}}}\xspace}

\begin{document}
\preprint{FERMILAB-PUB-19-272-ND}
\title{
    First measurement of neutrino oscillation parameters 
    using neutrinos and antineutrinos by NOvA
}
\newcommand{\ANL}{Argonne National Laboratory, Argonne, Illinois 60439, USA}
\newcommand{\ICS}{Institute of Computer Science, The Czech Academy of Sciences, 182 07 Prague, Czech Republic}
\newcommand{\IOP}{Institute of Physics, The Czech Academy of Sciences, 182 21 Prague, Czech Republic}
\newcommand{\Atlantico}{Universidad del Atlantico, Km. 7 antigua via a Puerto Colombia, Barranquilla, Colombia}
\newcommand{\BHU}{Department of Physics, Institute of Science, Banaras Hindu University, Varanasi, 221 005, India}
\newcommand{\UCLA}{Physics and Astronomy Department, UCLA, Box 951547, Los Angeles, California 90095-1547, USA}
\newcommand{\Caltech}{California Institute of Technology, Pasadena, California 91125, USA}
\newcommand{\Cochin}{Department of Physics, Cochin University of Science and Technology, Kochi 682 022, India}
\newcommand{\Charles}{Charles University, Faculty of Mathematics and Physics, Institute of Particle and Nuclear Physics, Prague, Czech Republic}
\newcommand{\Cincinnati}{Department of Physics, University of Cincinnati, Cincinnati, Ohio 45221, USA}
\newcommand{\CSU}{Department of Physics, Colorado State University, Fort Collins, CO 80523-1875, USA}
\newcommand{\CTU}{Czech Technical University in Prague, Brehova 7, 115 19 Prague 1, Czech Republic}
\newcommand{\Dallas}{Physics Department, University of Texas at Dallas, 800 W. Campbell Rd. Richardson, Texas 75083-0688, USA}
\newcommand{\DallasU}{University of Dallas, 1845 E Northgate Drive, Irving, Texas 75062 USA}
\newcommand{\Delhi}{Department of Physics and Astrophysics, University of Delhi, Delhi 110007, India}
\newcommand{\JINR}{Joint Institute for Nuclear Research, Dubna, Moscow region 141980, Russia}
\newcommand{\FNAL}{Fermi National Accelerator Laboratory, Batavia, Illinois 60510, USA}
\newcommand{\UFG}{Instituto de F\'{i}sica, Universidade Federal de Goi\'{a}s, Goi\^{a}nia, Goi\'{a}s, 74690-900, Brazil}
\newcommand{\Guwahati}{Department of Physics, IIT Guwahati, Guwahati, 781 039, India}
\newcommand{\Harvard}{Department of Physics, Harvard University, Cambridge, Massachusetts 02138, USA}
\newcommand{\Houston}{Department of Physics, University of Houston, Houston, Texas 77204, USA}
\newcommand{\IHyderabad}{Department of Physics, IIT Hyderabad, Hyderabad, 502 205, India}
\newcommand{\Hyderabad}{School of Physics, University of Hyderabad, Hyderabad, 500 046, India}
\newcommand{\IIT}{Department of Physics, Illinois Institute of Technology, Chicago IL 60616, USA}
\newcommand{\Indiana}{Indiana University, Bloomington, Indiana 47405, USA}
\newcommand{\INR}{Inst. for Nuclear Research of Russia, Academy of Sciences 7a, 60th October Anniversary prospect, Moscow 117312, Russia}
\newcommand{\Iowa}{Department of Physics and Astronomy, Iowa State University, Ames, Iowa 50011, USA}
\newcommand{\Irvine}{Department of Physics and Astronomy, University of California at Irvine, Irvine, California 92697, USA}
\newcommand{\Jammu}{Department of Physics and Electronics, University of Jammu, Jammu Tawi, 180 006, Jammu and Kashmir, India}
\newcommand{\Lebedev}{Nuclear Physics and Astrophysics Division, Lebedev Physical Institute, Leninsky Prospect 53, 119991 Moscow, Russia}
\newcommand{\MSU}{Department of Physics and Astronomy, Michigan State University, East Lansing, Michigan 48824, USA}
\newcommand{\Crookston}{Math, Science and Technology Department, University of Minnesota Crookston, Crookston, Minnesota 56716, USA}
\newcommand{\Duluth}{Department of Physics and Astronomy, University of Minnesota Duluth, Duluth, Minnesota 55812, USA}
\newcommand{\Minnesota}{School of Physics and Astronomy, University of Minnesota Twin Cities, Minneapolis, Minnesota 55455, USA}
\newcommand{\Oxford}{Subdepartment of Particle Physics, University of Oxford, Oxford OX1 3RH, United Kingdom}
\newcommand{\Panjab}{Department of Physics, Panjab University, Chandigarh, 160 014, India}
\newcommand{\Pitt}{Department of Physics, University of Pittsburgh, Pittsburgh, Pennsylvania 15260, USA}
\newcommand{\RAL}{Rutherford Appleton Laboratory, Science and Technology Facilities Council, Didcot, OX11 0QX, United Kingdom}
\newcommand{\SAlabama}{Department of Physics, University of South Alabama, Mobile, Alabama 36688, USA} 
\newcommand{\Carolina}{Department of Physics and Astronomy, University of South Carolina, Columbia, South Carolina 29208, USA}
\newcommand{\SDakota}{South Dakota School of Mines and Technology, Rapid City, South Dakota 57701, USA}
\newcommand{\SMU}{Department of Physics, Southern Methodist University, Dallas, Texas 75275, USA}
\newcommand{\Stanford}{Department of Physics, Stanford University, Stanford, California 94305, USA}
\newcommand{\Sussex}{Department of Physics and Astronomy, University of Sussex, Falmer, Brighton BN1 9QH, United Kingdom}
\newcommand{\Syracuse}{Department of Physics, Syracuse University, Syracuse NY 13210, USA}
\newcommand{\Tennessee}{Department of Physics and Astronomy, University of Tennessee, Knoxville, Tennessee 37996, USA}
\newcommand{\Texas}{Department of Physics, University of Texas at Austin, Austin, Texas 78712, USA}
\newcommand{\Tufts}{Department of Physics and Astronomy, Tufts University, Medford, Massachusetts 02155, USA}
\newcommand{\UCL}{Physics and Astronomy Dept., University College London, Gower Street, London WC1E 6BT, United Kingdom}
\newcommand{\Virginia}{Department of Physics, University of Virginia, Charlottesville, Virginia 22904, USA}
\newcommand{\WSU}{Department of Mathematics, Statistics, and Physics, Wichita State University, Wichita, Kansas 67206, USA}
\newcommand{\WandM}{Department of Physics, College of William \& Mary, Williamsburg, Virginia 23187, USA}
\newcommand{\Wisconsin}{Department of Physics, University of Wisconsin-Madison, Madison, Wisconsin 53706, USA}
\newcommand{\deceased}{Deceased.}
\affiliation{\ANL}
\affiliation{\Atlantico}
\affiliation{\BHU}
\affiliation{\Caltech}
\affiliation{\Charles}
\affiliation{\Cincinnati}
\affiliation{\Cochin}
\affiliation{\CSU}
\affiliation{\CTU}
\affiliation{\DallasU}
\affiliation{\Delhi}
\affiliation{\FNAL}
\affiliation{\UFG}
\affiliation{\Guwahati}
\affiliation{\Harvard}
\affiliation{\Houston}
\affiliation{\Hyderabad}
\affiliation{\IHyderabad}
\affiliation{\Indiana}
\affiliation{\ICS}
\affiliation{\IIT}
\affiliation{\INR}
\affiliation{\IOP}
\affiliation{\Iowa}
\affiliation{\Irvine}
\affiliation{\Jammu}
\affiliation{\JINR}
\affiliation{\Lebedev}
\affiliation{\MSU}
\affiliation{\Duluth}
\affiliation{\Minnesota}
\affiliation{\Panjab}
\affiliation{\Pitt}
\affiliation{\SAlabama}
\affiliation{\Carolina}
\affiliation{\SDakota}
\affiliation{\SMU}
\affiliation{\Stanford}
\affiliation{\Sussex}
\affiliation{\Syracuse}
\affiliation{\Tennessee}
\affiliation{\Texas}
\affiliation{\Tufts}
\affiliation{\UCL}
\affiliation{\Virginia}
\affiliation{\WSU}
\affiliation{\WandM}
\affiliation{\Wisconsin}
\author{M.~A.~Acero}
\affiliation{\Atlantico}

\author{P.~Adamson}
\affiliation{\FNAL}

\author{L.~Aliaga}
\affiliation{\FNAL}

\author{T.~Alion}
\affiliation{\Sussex}

\author{V.~Allakhverdian}
\affiliation{\JINR}

\author{S.~Altakarli}
\affiliation{\WSU}

\author{N.~Anfimov}
\affiliation{\JINR}

\author{A.~Antoshkin}
\affiliation{\JINR}

\affiliation{\Minnesota}

\author{A.~Aurisano}
\affiliation{\Cincinnati}

\author{A.~Back}
\affiliation{\Iowa}

% ...

\collaboration{The NOvA Collaboration}
\noaffiliation

\begin{abstract}

The NOvA experiment has made a $4.4\sigma$-significant observation
of $\antinue$ appearance in a 2~GeV $\antinumu$ beam at a distance of 810~km.
Using $12.33\times10^{20}$ protons on target delivered to the Fermilab NuMI 
neutrino beamline, the experiment recorded 27 $\numubar \rightarrow \nuebar$ 
candidates with a background of 10.3 and 102 
$\numubar \rightarrow \numubar$ candidates.
This new antineutrino data is combined with neutrino data to measure the
oscillation parameters $|\Delta m^2_{32}| = 2.48^{+0.11}_{-0.06}\times
10^{-3}$~eV$^2/c^4$, $\sin^2 \theta_{23} = 0.56^{+0.04}_{-0.03}$ in the normal
neutrino mass hierarchy and upper octant and excludes most values near 
$\delta_{\rm CP}=\pi/2$ for the inverted mass hierarchy by more 
than 3$\sigma$. The data favor the normal neutrino mass hierarchy by
1.9$\sigma$ and $\theta_{23}$ values in the upper octant by 1.6$\sigma$.

\end{abstract}

\maketitle

The observations of neutrino oscillations by many
experiments~\cite{Fukuda:1998mi,Fukuda:2002pe,Ahmad:2002jz,
Eguchi:2002dm,Michael:2006rx,Abe:2011sj,Abe:2011fz,An:2012eh,Ahn:2012nd}
are well described by the mixing of three neutrino mass eigenstates
$\nu_1$, $\nu_2$, and $\nu_3$ with the flavor eigenstates $\nu_e$,
$\nu_\mu$, and $\nu_\tau$.  The mixing is parameterized by a unitary
matrix, $U_{\rm PMNS}$, which depends on three angles and a phase,
\deltacp, that may break Charge-Parity (CP) symmetry.  The
oscillation frequencies are proportional to the neutrino mass
splittings, $\Delta m^2_{21} \equiv m_2^2 - m_1^2 \simeq
\SI{7.5e-5}{eV^2/{\it c}^4}$ and $|\Delta m^2_{32}| \simeq
\SI{2.5e-3}{eV^2/{\it c}^4}$, and the angles are known to be large:
$\theta_{12} \simeq 34^\circ$, $\theta_{13} \simeq 8^\circ$,
$\theta_{23} \simeq 45^\circ$~\cite{Patrignani:2016xqp}; 
$\deltacp$, however, is largely unknown.

Within this framework, several questions remain unanswered.  The angle
$\theta_{23}$ produces nearly maximal mixing but has large uncertainties.
If maximal, it would introduce an unexplained $\mu - \tau$ symmetry;
should it differ from $45^\circ$, its octant would determine whether
$\nu_\tau$ or $\nu_\mu$ couples more strongly to $\nu_3$.
Furthermore, while it is known that the two independent mass splittings 
differ by a factor of 30, the sign of the larger splitting is unknown.  
The $\nu_1$ and $\nu_2$ states that contribute most to the $\nu_e$ state
could be lighter (``normal hierarchy'', NH) or heavier (``inverted
hierarchy'', IH) than the $\nu_3$ state.
This question has important implications for models of 
neutrino mass~\cite{Mohapatra:2006gs,Nunokawa:2007qh,Altarelli:2010gt,King:2015aea,Petcov:2017ggy} 
and for the study of the 
Dirac vs.\ Majorana nature of the neutrino~\cite{Pascoli:2002xq,Bahcall:2004ip}.  
Additionally, neutrino mixing may be a source of CP violation 
if $\sin\delta_{\rm CP}$ is non-zero.

These questions can be addressed by the measurement of $\numu
\rightarrow \numu$, $\numubar \rightarrow \numubar$, $\numu
\rightarrow \nue$, and $\numubar \rightarrow \nuebar$ oscillations
in matter over baselines $L$ of order $(100 - 1000)\,\textrm{km}$,
with neutrino energies $E{\rm [GeV]} \simeq L{\rm [km]} \cdot |\Delta
m_{32}^2 {\rm[eV}^2/c^4{\rm ]}|$.  Several long-baseline experiments
have reported observations of $\numu \rightarrow
\numu$~\cite{Ahn:2006zza,Adamson:2014vgd,Abe:2018wpn,NOvA:2018gge},
$\numu \rightarrow
\nue$~\cite{Adamson:2014vgd,Abe:2018wpn,NOvA:2018gge}, and $\numubar
\rightarrow \numubar$~\cite{Adamson:2014vgd,Abe:2018wpn}, but a
statistically significant observation of $\numubar \rightarrow
\nuebar$ has not previously been made.  This report combines the first
antineutrino measurements using the NOvA detectors 
with the neutrino data reported in Ref.~\cite{NOvA:2018gge} 
in a reoptimized analysis yielding 
a new determination of the oscillation
parameters $|\Delta m^2_{32}|$, $\sin^2 \theta_{23}$, $\delta_{\rm CP}$,
and the neutrino mass hierarchy.

The NOvA experiment measures oscillations by comparing the
energy spectra of neutrino interactions in two detectors 
placed in the Fermilab NuMI beam~\cite{Adamson:2015dkw} at 
distances of \SI{1}{km} (Near Detector,
ND) and \SI{810}{km} (Far Detector, FD) from the production target.
The \SI{14}{kton} FD measures $\SI{15}{m} \times \SI{15}{m} \times
\SI{60}{m}$ while the \SI{290}{ton} ND consists of a $\SI{3.8}{m}
\times \SI{3.8}{m} \times \SI{12.8}{m}$ main detector followed by a
muon range stack.  Both detectors use liquid
scintillator~\cite{Mufson:2015kga} contained in PVC cells that are
$\SI{6.6}{cm} \times \SI{3.9}{cm}$ (0.15 radiation lengths $\times$
0.45 Moli\`{e}re radii) in cross section and span the height and width
of the detectors in planes of alternating vertical and horizontal orientation.  The
ND is located \SI{100}{m} underground. The FD operates on the surface
with modest shielding resulting in \SI{130}{kHz} of cosmic-ray
activity.  The detectors are located \SI{14.6}{mrad} off the beam axis
where the neutrino energy spectrum peaks at \SI{2}{GeV}. Magnetic focusing 
horns in the beamline charge-select neutrino parents giving 96\% (83\%) pure 
\numu (\numubar) event
samples between 1 and \SI{5}{GeV}. Most contamination is wrong-sign
($\bar\nu$ in the $\nu$ beam, or vice versa) with $<1\%$ $\nue +
\nuebar$ contamination.

This Letter reports data from an antineutrino beam run spanning from
June 29, 2016 to February 26, 2019, with an exposure of $12.33 \times
10^{20}$~protons-on-target (POT) delivered during \SI{317.0}{s} of
beam-on time, combined with the previously reported~\cite{NOvA:2018gge}
neutrino beam exposure of $8.85 \times 10^{20}$~POT and \SI{438.2}{s}.
During these periods, the proton source achieved a peak 
hourly-averaged power of \SI{742}{kW}.

The flux of neutrinos delivered to the detectors is calculated using a
simulation of the production and transport of particles through the
beamline components~\cite{Adamson:2015dkw,Agostinelli:2002hh} and
reweighted~\cite{Aliaga:2016oaz} to incorporate external measurements
of hadron production and 
interactions~\cite{Paley:2014rpb,Alt:2006fr,Abgrall:2011ae,Barton:1982dg,
Seun:2007zz,Tinti:2010zz,Lebedev:2007zz,Baatar:2012fua,Skubic:1978fi,
Denisov:1973zv,Carroll:1978hc,Abe:2012av,Gaisser:1975et,Cronin:1957zz,
Allaby:1969de,Longo:1962zz,Bobchenko:1979hp,Fedorov:1977an,Abrams:1969jm}.
Neutrino interactions in the detector are simulated using the
\textsc{genie} event generator~\cite{Andreopoulos:2009rq}.  The cross
section model has been tuned to improve agreement with external measurements
and ND data, reducing uncertainties in the extrapolation of
measurements in the ND to the FD.  As in Ref.~\cite{NOvA:2018gge}, we
set $M_A$ in the quasielastic dipole form factor to \SI{1.04}{GeV/{\it
c}^2}~\cite{Meyer:2016oeg} and use corrections to the charged-current~(CC)
quasielastic cross section derived from the random phase
approximation~\cite{Nieves:2004wx,Gran:2017psn}.  In this analysis, we
also apply this effect to baryon resonance production as a placeholder
for the unknown nuclear effect that produces a suppression observed
at low four-momentum transfer in our and other
measurements~\cite{Adamson:2014pgc,AguilarArevalo:2010bm,
McGivern:2016bwh,Altinok:2017xua}.
Additionally, we increase the rate of deep-inelastic scattering with
hadronic mass $W>\SI{1.7}{GeV/{\it c}^2}$ by 10\% to match our
observed rates of short track-length $\numu$~CC events.  We model
multi-nucleon ejection interactions following
Ref.~\cite{Katori:2013eoa} and adjust the rates in bins of energy
transfer, $q_0$, and 3-momentum transfer, $|\vec{q}|$, for \numu and
\numubar separately to maximize agreement in the ND.  The calculation
of the \nue and \nuebar rates uses these same models.

The energy depositions of final-state particles are simulated with
\textsc{geant4}~\cite{Agostinelli:2002hh} and input to a
custom simulation of the production of, and the detector response to,
scintillation and Cherenkov light~\cite{Aurisano:2015oxj}.  The
absolute energy scale of the detectors is calibrated to within 
$\pm 5\%$ using the minimum ionizing portion of cosmic-ray muon 
tracks that stop in the detectors.

Cells with activity above threshold (hits) are grouped based on their
proximity in space and time to produce candidate neutrino events.
Events are assigned a vertex, and clusters are formed from hits likely
to be associated with particles produced there~\cite{Baird:2015pgm}.
These clusters are categorized as electromagnetic
or hadronic in origin using a convolutional neural network
(CNN)~\cite{Psihas:2018czu}.  Hits forming tracks are identified as
muons by combining information on the track length, $dE/dx$, vertex
activity, and scattering into a single particle identification (PID)
score~\cite{Raddatz:2016iui}.  The same reconstruction algorithms are
applied to events from data and simulation in both detectors.

The $\numu$ and $\numubar$ candidates are required to have a vertex 
inside the fiducial volume and no evidence of particles exiting the detector.
The $\nue$ and $\nuebar$ candidates are divided into a ``core'' sample which
satisfies these containment requirements, and a ``peripheral'' sample
which loosens these requirements for the most signal-like event
topologies.  A second CNN~\cite{Aurisano:2016jvx} serves as the primary
PID, classifying event topologies as $\nue$~CC, $\numu$~CC,
$\nutau$~CC, neutral-current (NC), or cosmic ray.  The network is
trained on simulated neutrino events and cosmic-ray data, 
separately for neutrino and antineutrino beam conditions.  It has an
improved architecture and higher rate of cosmic ray rejection over the
previous network~\cite{NOvA:2018gge}.  Events identified as $\numu$~CC
are further required to contain at least one track classified as a
muon.

Several requirements further reduce cosmic-ray backgrounds.  For the
\numu~CC sample, a boosted decision tree (BDT) algorithm based on
vertex position and muon-like track properties is used. Events in the
core \nue sample not aligned with the beam direction and that are
near the top of the detector are rejected.  Events characterized as
detached bremsstrahlung showers from cosmic tracks are also removed,
as are events whose topology is consistent with photons entering from
the detector north side where there is less shielding.  Events in the
\nue peripheral sample are tested against a BDT classifier using event
position and direction information to separate them from cosmic-ray 
topologies.

The selection of $\numu$ and $\numubar$~CC events is 31.2\% (33.9\%)
efficient relative to true interactions in the fiducial volume,
resulting in 98.6\% (98.8\%) pure samples at the FD
during neutrino (antineutrino) beam operation.
Both $\numu$ and $\numubar$ are counted as signal for the disappearance 
measurements. Selections against exiting particle tracks are the 
largest source of inefficiency.  The efficiency for selecting signal \nue~CC
($\nuebar$~CC) events is 62\% (67\%).  Purities for the signal \nue
(\nuebar) samples fall in the range 57--78\% (55--77\%)
depending on the impact of oscillations on the signal and wrong-sign
background levels.  These efficiencies and purities differ from those
quoted in Ref.~\cite{NOvA:2018gge} due to a reoptimization of the 
selection algorithms~\cite{Blackburn:2019ngc}.  The wrong-sign component
of the selected $\numu$ sample in the ND is calculated to be 
$2.8\pm0.3\%$ and $10.6\pm1.1\%$ 
for the neutrino and antineutrino beams. These 
fractions were found to be consistent with a data-driven estimate 
based on the rate of $\numu$~CC and NC interactions with associated 
detector activity indicative of neutron capture.

The incident neutrino energy is reconstructed from the measured
energies of the final-state lepton and recoil hadronic system.
The lepton energy is estimated from track length for muon candidates and
from calorimetric energy for electron candidates.  
The hadronic energy is estimated from the sum of the
calibrated hits not associated with the primary lepton. 
The neutrino energy resolution at the FD is 9.1\% (8.1\%) 
for \numu~CC (\numubar~CC) events and 10.7\% (8.8\%) for 
\nue~CC (\nuebar~CC) events.  The \numu and \numubar 
events with the lowest hadronic energy fraction give the
best energy resolution and lowest backgrounds, yielding the most
precise measurement of the oscillated spectral shape, so we 
analyzed the spectra separately in quartiles of this
variable~\cite{NOvA:2018gge}.

\begin{figure*}
    \includegraphics[width=1.78in]{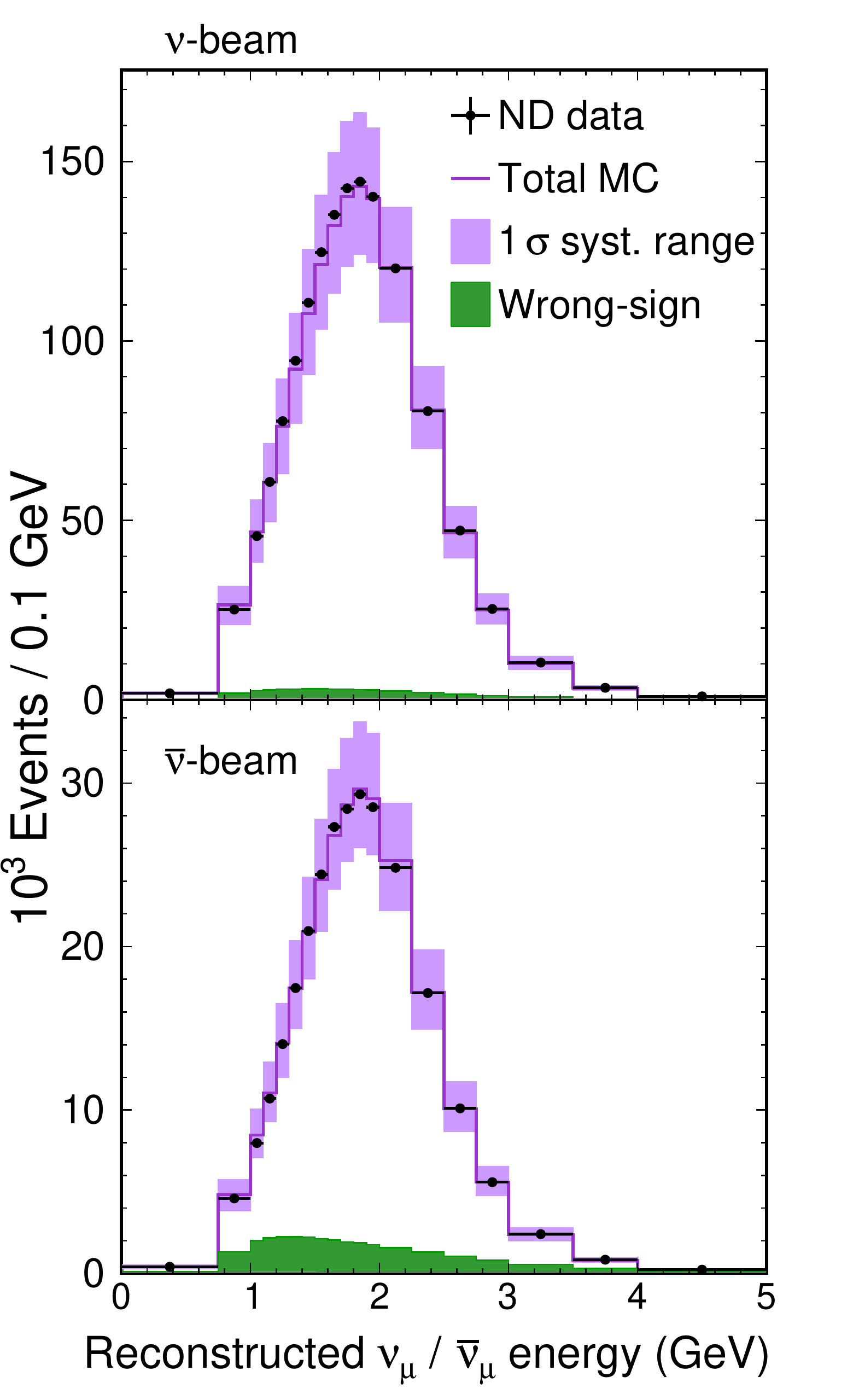}\includegraphics[width=1.78in]{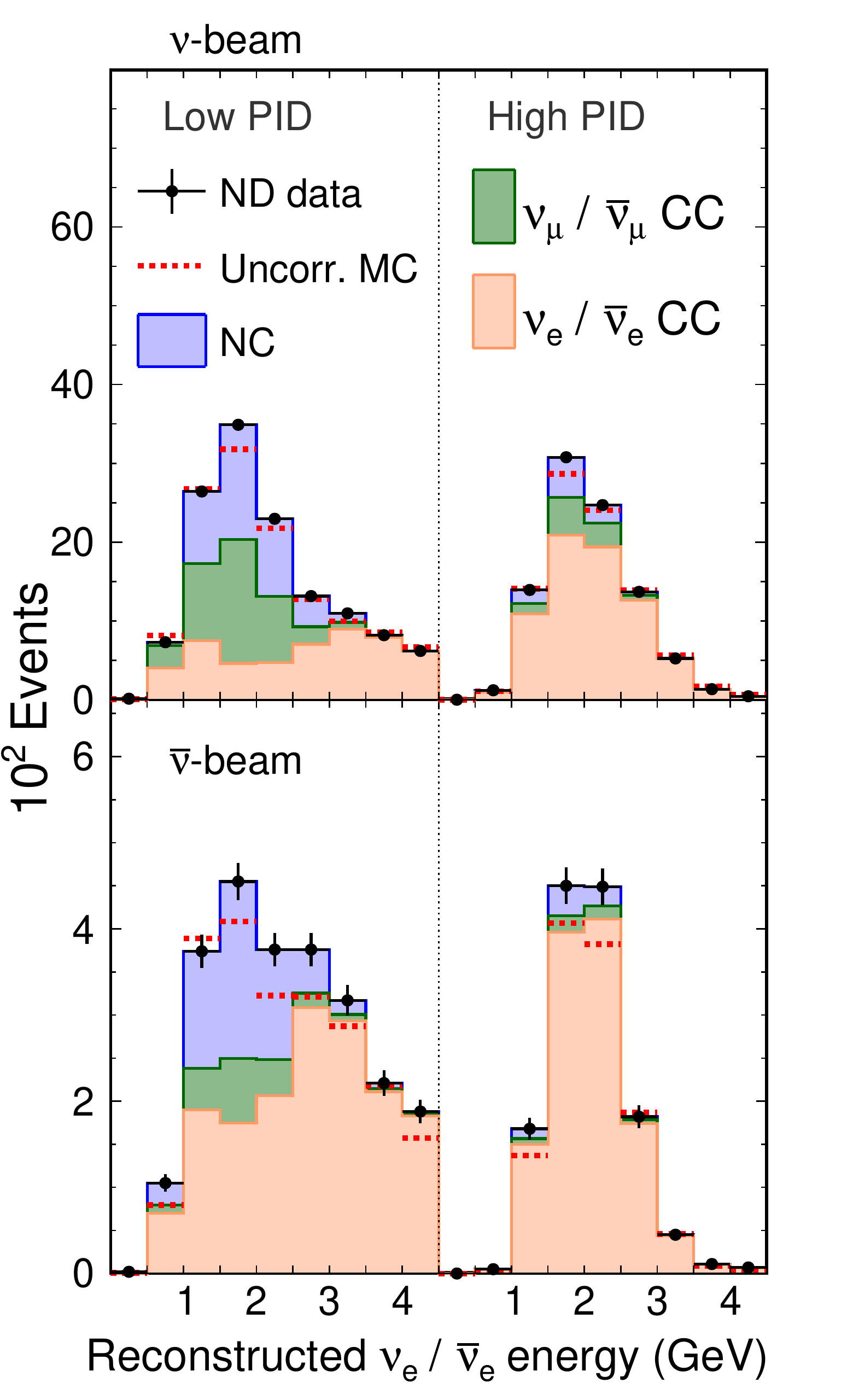}~
    \includegraphics[width=1.78in]{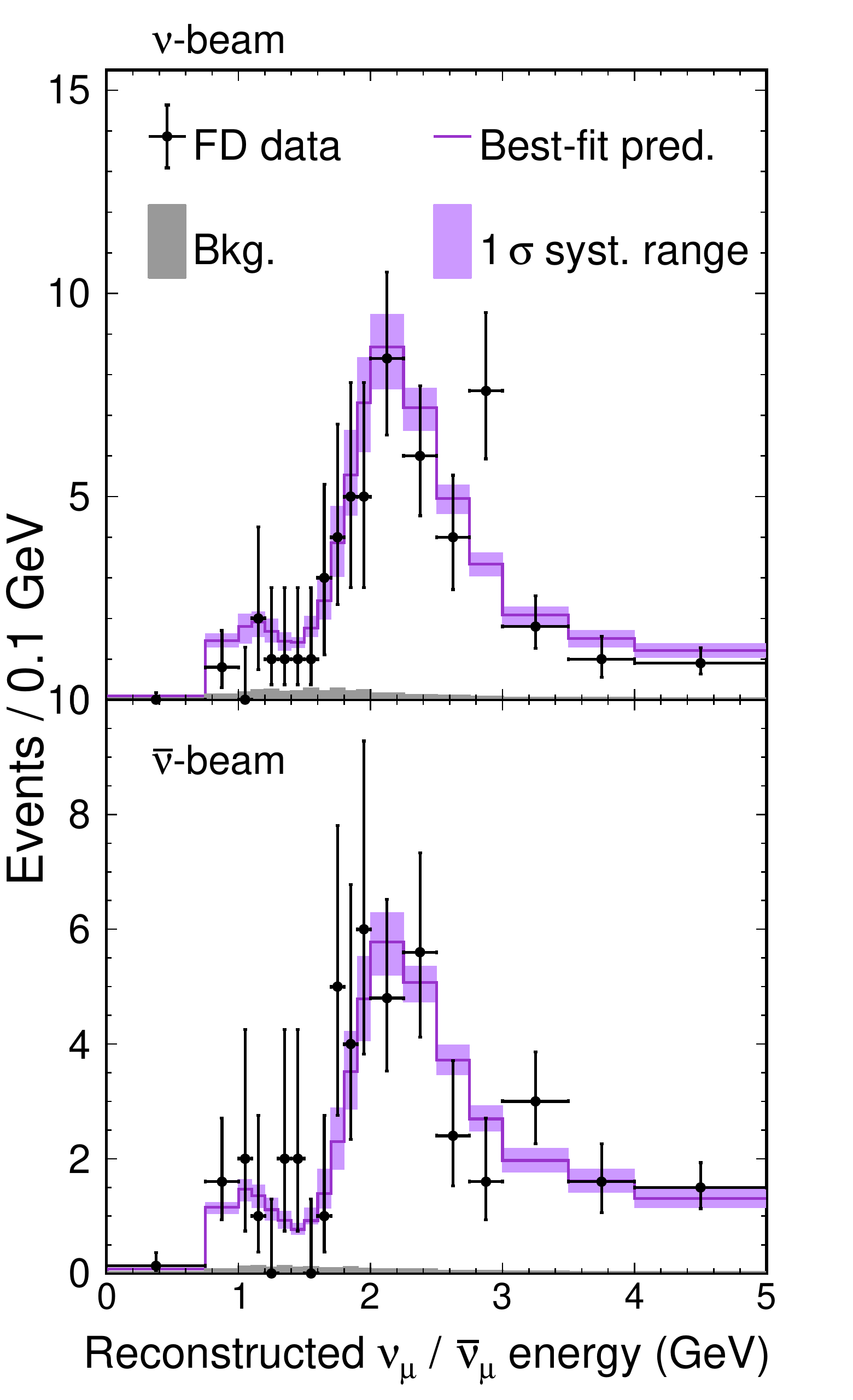}\includegraphics[width=1.78in]{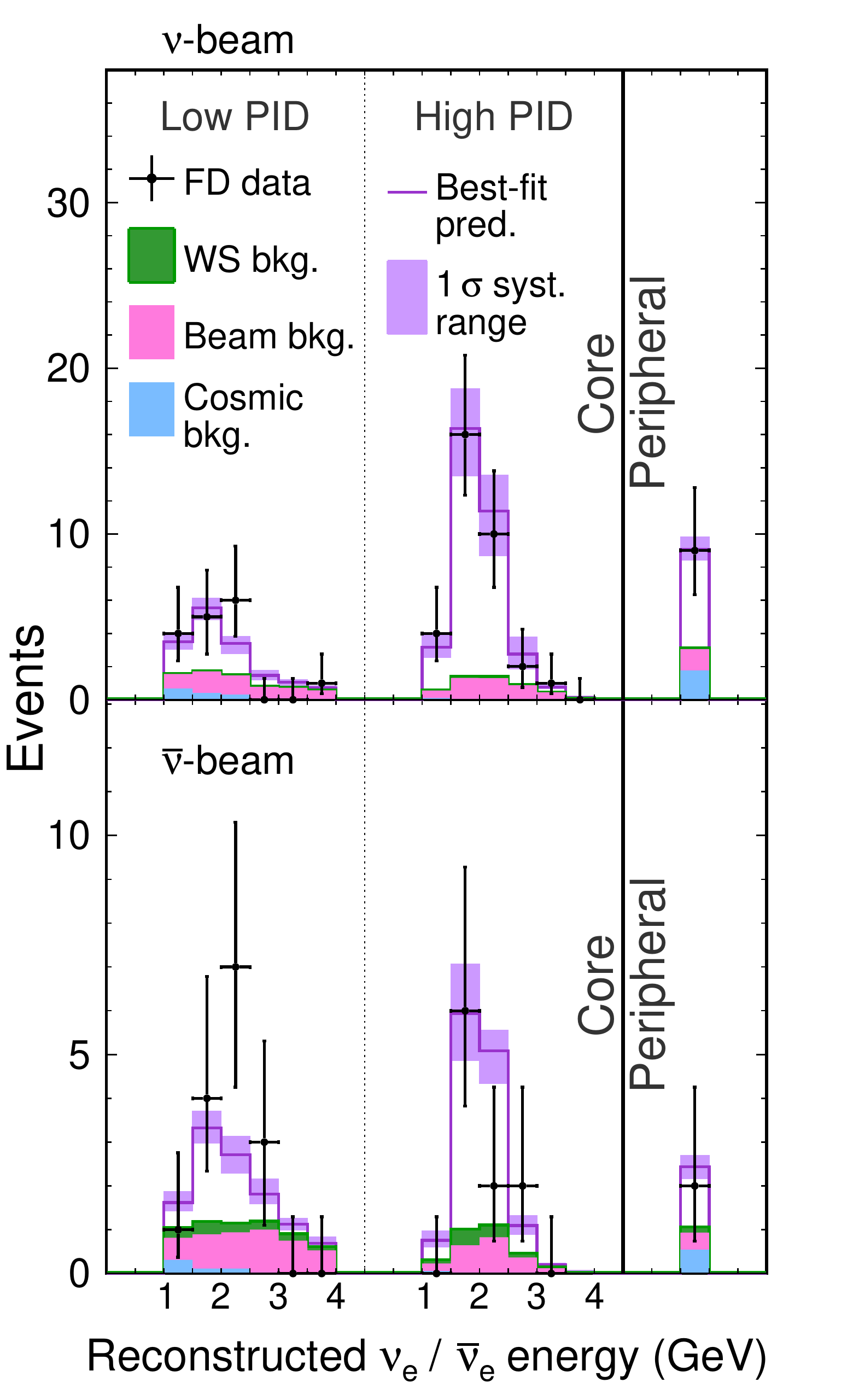}
\caption{
From left to right, 
the reconstructed neutrino energy spectra for the 
ND $\numu$~CC, ND $\nue$~CC,
FD $\numu$~CC, FD $\nue$~CC~\cite{supplemental-numu} with 
neutrino beam on the top and antineutrino beam on the bottom.
For the ND \numu~CC spectra, backgrounds aside from wrong-sign 
are negligible and not shown.
The \nue~CC spectra are split into a low and high purity sample, 
and the FD spectra shows 
counts in the ``peripheral'' sample. The dashed lines in the 
ND \nue spectra show the totals before data-driven corrections.
}
\label{fig:nd-fd-data-figure}
\end{figure*}

The energy spectra of the selected $\numu$~CC and $\nue$~CC
interactions in the ND during neutrino and antineutrino beam
operations are shown in Fig.~\ref{fig:nd-fd-data-figure}.  The selected ND
$\nue$ sample consists entirely of background sources for the $\nue$
appearance measurement, predominantly the intrinsic beam $\nue$
component, along with misidentified $\numu$~CC and NC
interactions.  We analyze the $\nue$ candidate energy spectra in two
bins of $\nue$ PID (``low'' and ``high'') to isolate a highly pure
sample of $\numu \rightarrow \nue$ and $\numubar \rightarrow \nuebar$
at the FD. In the ND, the high-PID sample is dominated by intrinsic
beam $\nue$.  A third bin containing the ``peripheral'' events is
added for the FD.

The \numu and \nue signal spectra at the FD are predicted for the
neutrino and antineutrino beams separately and are based on the
observed spectra of \numu candidate events in the ND.  The true neutrino
energy spectrum at the ND is estimated using the measured event rates
in bins of reconstructed energy and the energy distributions of simulated
events found to populate those bins.  
This true spectrum is corrected
for differences in flux and acceptance between the ND and FD, as well
as differences in the \numu and \nue cross sections;
oscillations are then applied to yield predictions for the true \numu and
\nue spectra at the FD.  These spectra are then transformed into
reconstructed energy using the underlying energy
distributions from simulated neutrino interactions in the FD.

The predicted background spectra at the FD are also primarily
data-driven.  Data collected out-of-time with the NuMI beam provide a
measurement of the rate of cosmic-ray backgrounds in the \numu and
\nue samples. Neutrino backgrounds calculated to populate the FD \nue
spectra are corrected based on the reconstructed \nue candidates at the
ND.  The procedure from Ref.~\cite{NOvA:2018gge} is followed to
determine corrections for each background component in the
neutrino-mode beam, while for the antineutrino-mode beam a single
scale factor is used.  The remaining backgrounds, which include
any misidentified neutrino events in the \numu samples and misidentified 
\nutau interactions in the \nue samples, make up less than 2\% of the FD
candidates and are taken directly from simulation.

To evaluate the impact of systematic uncertainties we recompute the
extrapolation from the ND to the FD varying the parameters used to
model the neutrino fluxes, neutrino cross sections, and the detector
response.  The procedure accounts for changes in the composition of the 
\nue background, and for impact on the transformation to and from 
true and reconstructed energies due to 
variations in the model parameters.
We parameterize each systematic variation and
compute its effect in each analysis bin.  These parameters are
included in the oscillation fit constrained within their estimated
uncertainties by penalty terms in the likelihood function.

\begin{table}[htb]
    \caption{Systematic uncertainties on the total predicted numbers of 
    signal and beam-related background events at the best fit point (see Table~\ref{tb:oscillation_parameters_table}) in the \nue selected 
    samples in the neutrino and antineutrino datasets.}
    \label{tb:systs-nue}
    \begin{tabular}{l c c c c }
        \hline \hline
                  & \nue Signal  & \nue Bkg. & \nuebar Signal & \nuebar Bkg. \\
                  Source & (\%)  & (\%) & (\%) & (\%) \\
        \hline
            Cross-sections & +4.7/-5.8 & +3.6/-3.4 & +3.2/-4.2 & +3.0/-2.9 \\
            Detector model & +3.7/-3.9 & +1.3/-0.8 & +0.6/-0.6 & +3.7/-2.6 \\
            ND/FD diffs.   & +3.4/-3.4 & +2.6/-2.9 & +4.3/-4.3 & +2.8/-2.8 \\
            Calibration    & +2.1/-3.2 & +3.5/-3.9 & +1.5/-1.7 & +2.9/-0.5 \\
            Others         & +1.6/-1.6 & +1.5/-1.5 & +1.4/-1.2 & +1.0/-1.0 \\
        \hline            
            Total             & +7.4/-8.5 & +5.6/-6.2 & +5.8/-6.4 & +6.3/-4.9 \\
        \hline \hline
    \end{tabular}
\end{table}

\begin{table}[htb]
    \caption{Systematic and statistical uncertainties on
    the oscillation parameters $\sin^2 \theta_{23}$, $\Delta m^2_{32}$, and $\delta_{\rm CP}$, evaluated at the 
    best fit point (see Table~\ref{tb:oscillation_parameters_table}).
    }
    \label{tb:systs-numu}
    \begin{tabular}{l c c c}
        \hline \hline
\rule{0pt}{9pt} & \snsq & $|\dmsq|$ & $\deltacp$\\
                  Source &
                  {\footnotesize ($\times 10^{-3}$)} &
                  {\footnotesize ($\times 10^{-5}~\mathrm{eV}^2/c^4$)} & 
                  {\footnotesize ($\pi$)} \\
        \hline
            Calibration      & +5.4 / -9.2  & +2.2 / -2.6 & +0.03 / -0.03 \\
            Neutron model    & +6.0 / -13.0 & +0.5 / -1.3 & +0.01 / -0.00 \\
            Cross-sections   & +4.1 / -7.7  & +1.0 / -1.1 & +0.06 / -0.07 \\
            $E_\mu$ scale    & +2.3 / -3.0  & +1.0 / -1.1 & +0.00 / -0.00 \\
            Detector model   & +1.9 / -3.2  & +0.4 / -0.5 & +0.05 / -0.05 \\
            Normalizations   & +1.3 / -2.7  & +0.1 / -0.2 & +0.02 / -0.03 \\
            ND/FD diffs.     & +1.0 / -4.0  & +0.2 / -0.2 & +0.06 / -0.07 \\
            Beam flux        & +0.4 / -0.8  & +0.1 / -0.1 & +0.00 / -0.00 \\
        \hline
            Total syst.      & +9.7 / -20   & +2.6 / -3.2 & +0.11 / -0.12\\
        \hline \hline
    \end{tabular}
\end{table}

The oscillation parameters that best fit the FD data are determined
through minimization of a Poisson negative log-likelihood, \LL,
considering three unconstrained parameters, \dmsq, \snsq, and \deltacp,
as well as 53 constrained parameters covering the other PMNS
oscillation parameters and the sources of systematic uncertainty 
summarized in Tables \ref{tb:systs-nue} and \ref{tb:systs-numu}.  The
two-detector design and extrapolation procedure significantly reduce
the effect of the $\simeq$10--20\% a priori uncertainties on
the beam flux and cross sections.  
The principal remaining uncertainties are 
neutrino cross sections,
the energy scale calibration,
the detector response to neutrons, 
and differences between the ND and FD
that cannot be corrected by extrapolation.

\begin{table}
\caption{
Event counts at the FD, both observed and predicted at the best fit point (see Table~\ref{tb:oscillation_parameters_table}).
}
\label{tb:event-count-table}
\begin{tabular}{lcccc}
	\hline
	\hline
           & \multicolumn{2}{c}{Neutrino beam} & \multicolumn{2}{c}{Antineutrino beam} \\
& {$\numu$~CC}  & {$\nue$~CC}  & {$\numubar$~CC} & {$\nuebar$~CC} \\
    \hline
        $\numu \rightarrow \numu$      &  112.5        & 0.7          & 24.0            & 0.1  \\
        $\numubar \rightarrow \numubar$&  7.2          & 0.0          & 70.0            & 0.1  \\
        $\numu    \rightarrow \nue$    &  0.1          & 44.3         & 0.0             & 2.2  \\
        $\numubar \rightarrow \nuebar$ &  0.0          & 0.6          & 0.0             & 16.6 \\
        Beam $\nue+\nuebar$            &  0.0          & 7.0          & 0.0             & 5.3  \\
        NC                             &  1.3          & 3.1          & 0.8             & 1.2  \\
        Cosmic                         &  2.1          & 3.3          & 0.8             & 1.1  \\
        Others                         &  0.7          & 0.4          & 0.6             & 0.3  \\
    \hline
\rule{0pt}{9pt}Signal            & 120$^{+10}_{-12}$ & 44.3$^{+3.5}_{-4.0}$ & 93.9$^{+8.1}_{-8.2}$ & 16.6$^{+0.9}_{-1.0}$ \\
\rule{0pt}{9pt}Background    & 4.2$^{+0.5}_{-0.6}$     & 15.0$^{+0.8}_{-0.9}$ & $2.2^{+0.4}_{-0.4}$ &  10.3$^{+0.6}_{-0.5}$  \\
    \hline
\rule{0pt}{9pt}Best fit                & 124          & 59.3       & 96.2            & 26.8 \\
               Observed                & 113          & 58         & 102              & 27   \\
    \hline
    \hline
    \end{tabular}
\end{table}

The selection criteria and techniques used in the analysis were
developed on simulated data prior to inspection of the FD data
distributions.  Figure~\ref{fig:nd-fd-data-figure} shows the energy
spectra of the $\numu$~CC, $\numubar$~CC, $\nue$~CC, and $\nuebar$~CC
candidates recorded at the FD overlaid on their oscillated best-fit
expectations.  Table~\ref{tb:event-count-table} summarizes the total
event counts and estimated compositions of the selected samples.  We
recorded 102 \numubar candidate events at the FD,
reflecting a significant suppression from the unoscillated
expectation of 476.  We find 27 $\numubar \rightarrow \nuebar$
candidate events with an estimated background of $10.3^{+0.6}_{-0.5}$,
a \SI{4.4}{\ensuremath{\sigma}} excess over 
the predicted background. This observation is the first evidence
of $\nuebar$ appearance in a $\numubar$ beam over a long baseline.
These new antineutrino data are analyzed together with
113~\numu and 58~$\numu \rightarrow \nue$
candidates from the previous data set.

\begin{table}
    \caption{
        Summary of oscillation parameters. 
        The top three are inputs to this analysis~\cite{Patrignani:2016xqp}, while the rest
        are the best fits for different choices of the mass hierarchy (NH, IH) and $\theta_{23}$ 
        octant (UO, LO), along with the significance (in units of $\sigma$) at which those
        combinations are disfavored.
        In addition to the region indicated, for NH, LO a small range of $\sin^2 \theta_{23}$
        $0.45-0.48$ is allowed at $1\sigma$~\cite{supplemental-contours}.
}
    \label{tb:oscillation_parameters_table}
    \begin{tabular}{r|cccc}
        \hline \hline
            \rule{0pt}{9pt} $\Delta m^2_{21}/(10^{-5}~{\rm eV}^2/c^4)$ & \multicolumn{4}{c}{$7.53 \pm 0.18$} \\
                            $\sin^2 \theta_{12}$                       & \multicolumn{4}{c}{$0.307^{+0.013}_{-0.012}$} \\
                            $\sin^2 \theta_{13}$                       & \multicolumn{4}{c}{$0.0210 \pm 0.0011$} \\
        \cline{2-5}
            \rule{0pt}{8pt}                                            & NH, UO                  & NH, LO        & IH, UO       & IH, LO \\
        \cline{2-5}
            \rule{0pt}{9pt} $\Delta m^2_{32}/(10^{-3}~{\rm eV}^2/c^4$) & $+2.48^{+0.11}_{-0.06}$ & $+2.47$       & $-2.54$      & $-2.53$ \\
                            $\sin^2 \theta_{23}$                       & $0.56^{+0.04}_{-0.03}$  & $0.48$        & $0.56$       & $0.47$ \\
            \rule{0pt}{9pt} $\delta_{\rm CP}/\pi$                      & $0.0^{+1.3}_{-0.4}$     & $1.9$         & $1.5$        & $1.4$ \\
        \cline{2-5}
                                                                       & -                       & $+1.6\sigma$  & $+1.8\sigma$ & $+2.0\sigma$ \\
        \hline \hline
    \end{tabular}
\end{table}

Table~\ref{tb:oscillation_parameters_table} shows the overall best-fit
parameters, as well as the best fits for each choice of $\theta_{23}$
octant and hierarchy. The best-fit point is found for the normal
hierarchy with $\theta_{23}$ in the upper octant where \LL = 157.1 for
175 degrees of freedom (goodness-of-fit $p=0.91$ from simulated
experiments). The measured values of $\theta_{23}$ and 
$\Delta m^2_{32}$ are consistent with the previous NOvA 
measurement~\cite{NOvA:2018gge} that used only neutrino data, 
and are consistent with maximal mixing within $1.2\sigma$. 

\begin{figure}[htb]
    \includegraphics[width=3.5in]{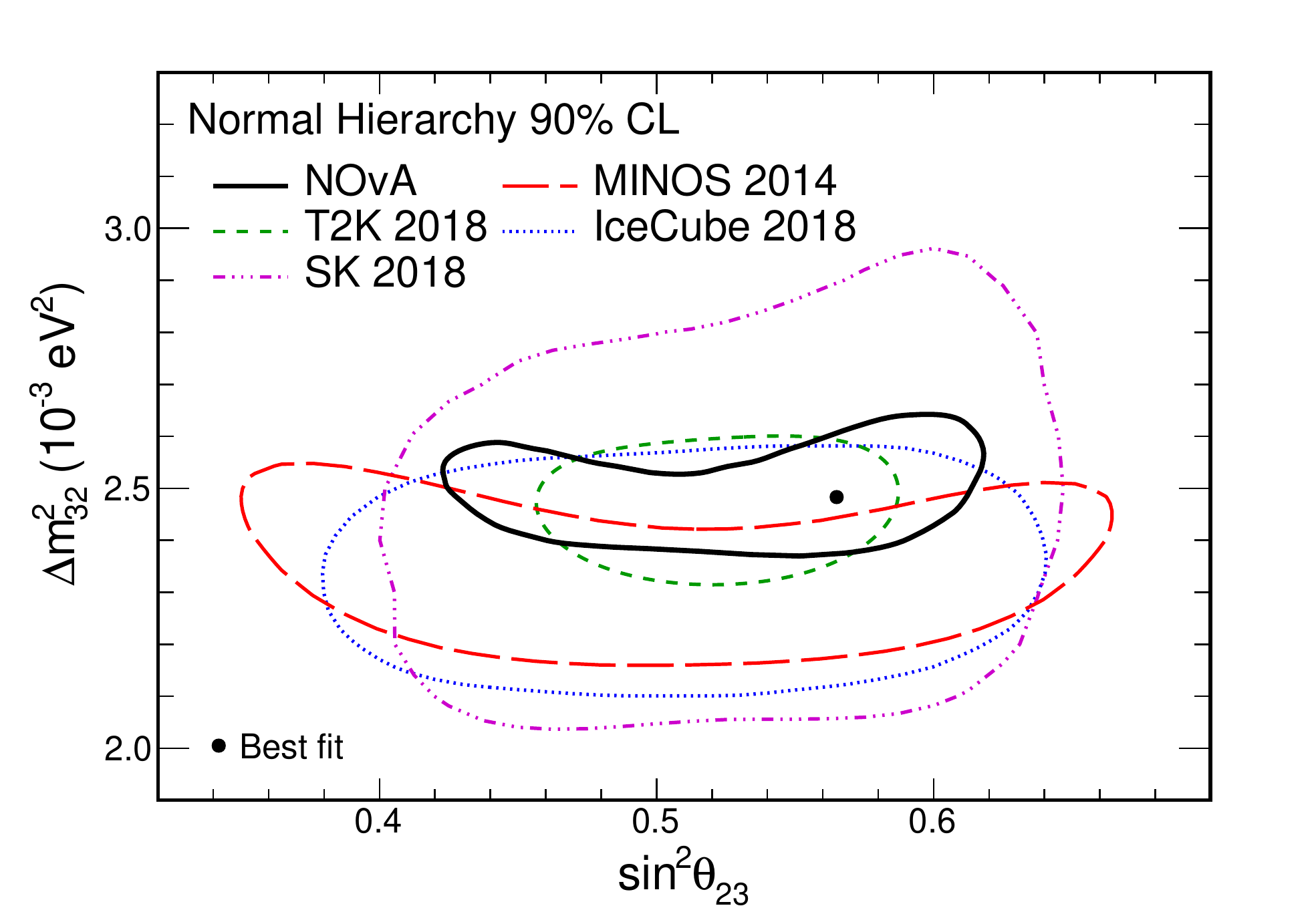}
    \caption{The 90\% confidence level region for \dmsq and \snsq, 
    with best-fit point shown as a black marker~\cite{supplemental-contours}, 
    overlaid on contours from other experiments~\cite{Abe:2018wpn,Abe:2017aap,Adamson:2014vgd,Aartsen:2017nmd}. 
}
\label{fig:dms32-sstt23-contour}
\end{figure}
\begin{figure}[htb]
    \includegraphics[width=3.5in]{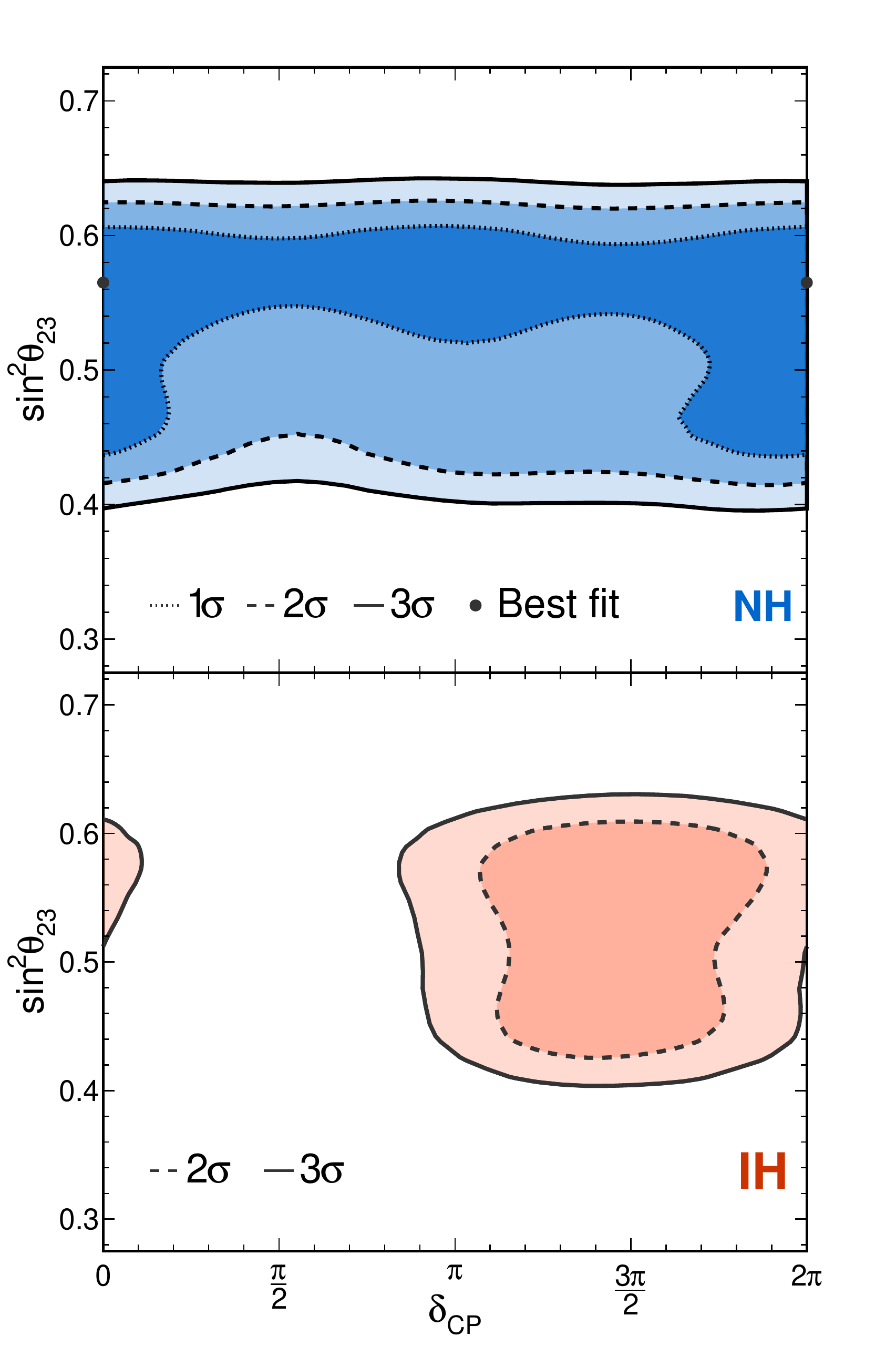}
    \caption{The $1\sigma$, $2\sigma$, and $3\sigma$ contours in 
    \snsq vs. \deltacp in the normal hierarchy (NH, top panel) 
    and inverted hierarchy (IH, bottom panel)~\cite{supplemental-dcp}. 
    The best-fit point is shown by a black marker.
    }
    \label{fig:sstt23-dcp-islands}
\end{figure}
Confidence intervals for the oscillation parameters are determined
using the unified approach~\cite{Feldman:1997qc}, as detailed in
Ref.~\cite{Sousa2019aa}.  Figure~\ref{fig:dms32-sstt23-contour}
compares the 90\% confidence level contours in $\dmsq$ and $\snsq$ 
with those of other other experiments~\cite{Adamson:2014vgd,Aartsen:2017nmd,Abe:2017aap,Abe:2018wpn}.
Figure~\ref{fig:sstt23-dcp-islands} shows the allowed regions in \snsq
and \deltacp. 
These results exclude most values near $\delta_{\rm CP}=\pi/2$ 
in the inverted mass hierarchy by more than 3$\sigma$;
specifically the intervals between 
\numrange{-0.04}{0.97}$\pi$
in the lower $\theta_{23}$ octant and 
\numrange{0.04}{0.91}$\pi$
in the upper octant.
The data prefer the normal hierarchy with a significance
of $1.9\sigma$ ($p = 0.057$, $CL_s = 0.091$~\cite{Read:2002hq}) 
and the upper $\theta_{23}$ octant with a
significance of $1.6\sigma$ ($p = 0.11$), 
profiling over all other parameter choices.

\begin{acknowledgments}
We are grateful to Stephen Parke (FNAL) for useful discussions.
This document was prepared by the NOvA collaboration using the
resources of the Fermi National Accelerator Laboratory 
(Fermilab), a U.S. Department of Energy, Office of Science, 
HEP User Facility. Fermilab is managed by Fermi Research Alliance, 
LLC (FRA), acting under Contract No. DE-AC02-07CH11359. This work 
was supported by the U.S. Department of Energy; the
U.S. National Science Foundation; the Department of Science and
Technology, India; the European Research Council; the MSMT CR, GA UK,
Czech Republic; the RAS, RFBR, RMES, RSF, and BASIS Foundation,
Russia; CNPq and FAPEG, Brazil; STFC, and the Royal Society, United
Kingdom; and the state and University of Minnesota. This work used
resources of the National Energy Research Scientific Computing Center
(NERSC), a U.S. Department of Energy Office of Science User Facility
operated under Contract No. DE-AC02-05CH11231.  We are grateful for the 
contributions of the staffs of the University of Minnesota at 
the Ash River Laboratory and of Fermilab.

\end{acknowledgments}

\bibliography{refs}

\end{document}